\documentclass[conference]{IEEEtran}
\IEEEoverridecommandlockouts
\usepackage{cite}
\usepackage{amsmath,amssymb,amsfonts}
\usepackage{graphicx}
\usepackage{textcomp}
\usepackage{xcolor}

\usepackage{algorithm}
\usepackage{algpseudocode}

\def\BibTeX{{\rm B\kern-.05em{\sc i\kern-.025em b}\kern-.08em
    T\kern-.1667em\lower.7ex\hbox{E}\kern-.125emX}}
\newcommand*\diff{\mathop{}\!\mathrm{d}}

\begin{document}

\title{On the Benefits of Anticipating Load Imbalance for Performance Optimization of Parallel Applications}

\author{
    \IEEEauthorblockN{Anthony Boulmier\IEEEauthorrefmark{1}, Franck Raynaud\IEEEauthorrefmark{1}, Nabil Abdennadher\IEEEauthorrefmark{2}, Bastien Chopard\IEEEauthorrefmark{1}}
    \IEEEauthorblockA{\IEEEauthorrefmark{1}\textit{Departement of Computer Science}\\
    \textit{University of Geneva}\\
    Geneva, Switzerland\\
    \{anthony.boulmier, franck.raynaud, bastien.chopard\}@unige.ch}
    \IEEEauthorblockA{\IEEEauthorrefmark{2}\textit{Departement of Computer Science} \\
    \textit{University of Applied Sciences Western Switzerland}\\
    Geneva, Switzerland\\
    nabil.abdennadher@hesge.ch}
}
\IEEEpubid{\makebox[\columnwidth]{
978-1-7281-4734-5/19/\$31.00~\copyright2019 IEEE \hfill} \hspace{\columnsep}\makebox[\columnwidth]{ }}

\maketitle

\IEEEpubidadjcol

\begin{abstract}
In parallel iterative applications, computational efficiency is essential for addressing large problems. Load imbalance is one of the major performance degradation factors of parallel applications. Therefore, distributing, cleverly, and as evenly as possible, the workload among processing elements (PE) maximizes application performance. So far, the standard load balancing method consists in distributing the workload evenly between PEs and, when load imbalance appears, redistributing the extra load from overloaded PEs to underloaded PEs. However, this does not anticipate the load imbalance growth that may continue during the next iterations. In this paper, we present a first step toward a novel philosophy of load balancing that unloads the PEs that will be overloaded in the near future to let the application rebalance itself via its own dynamics. 
Herein, we present a formal definition of our new approach using a simple mathematical model and discuss its advantages compared to the standard load balancing method. In addition to the theoretical study, we apply our method to an application that reproduces the computation of a fluid model with non-uniform erosion. The performance validates the benefit of anticipating load imbalance. We observed up to $16\%$ performance improvement compared to the standard load balancing method.
\end{abstract}


\section{INTRODUCTION}

Parallel iterative applications are widely used in quantum physics, computational fluid dynamics, molecular dynamics, big data analytics, and others. In particular, we are interested in iterative applications that are executed on distributed memory HPC infrastructures.
Among them, many are subject to important performance degradation factors that need to be minimized for addressing large problems. One of the major performance degradation factor is load imbalance among processing elements~(PE). Most of the time, load imbalance arises from problem, algorithmic, and systemic characteristics~\cite{Boulmier2017}. To mitigate load imbalance, load balancing (LB) techniques have been developed over the years. All those techniques use the same strategy: partitioning the problem space into almost equal sub-domains and mapping them onto their most suitable PE in order to minimize the wall clock time of the application. In this paper, we refer to this strategy as the \textit{standard method} of LB. Popular partitioning techniques include recursive coordinate bisection, recursive inertial bisection, space filling curves, graph partitioning algorithms, and others~\cite{Devine2005}. In irregular dynamic applications, where the distribution of the workload among the computational domain changes over time, multiple calls to the LB technique are needed to react to the load imbalance growth. Indeed, LB techniques are not able to anticipate load imbalance and thus load balancing is repeatedly required.  Moreover, deciding the optimal time to activate the LB technique is an important factor of performance optimization when dealing with irregular dynamic applications. When addressing large problems involving thousands of PEs, LB techniques have a non-negligible computational cost, which is difficult to predict~\cite{Calotoiu2018, Lee2007}. Therefore, they must be called only when needed. So far, the best way to do so consists in calling the load balancer when the load imbalance cost is greater than the LB cost~\cite{Pearce2012, H.MenonandN.JainandG.ZhengandL.Kale2012}. This requires a cost model of the application as well as one of the load balancer. Previous works modeled parallel applications with the standard LB method allowing efficient adaptive LB calls~\cite{Pearce2012, H.MenonandN.JainandG.ZhengandL.Kale2012}. A recent work proposed by Zhai et al.~\cite{Zhai2018} improved the work presented by Menon et al.~\cite{H.MenonandN.JainandG.ZhengandL.Kale2012} making the LB calls more flexible. However, these works only react to the load imbalance instead of anticipating its growth, wasting a lot of computing resources especially as we are going toward exascale computing. 


In this paper, we present the first stage in the direction of a new LB approach that anticipates the load imbalance growth and uses this information to partition the workload among the PEs for iterative applications running on distributed memory HPC infrastructures. 
The main contributions of this paper are as follows: (i) we use the workload increase rate (WIR) of the PEs to predict the load imbalance growth in order to drive the domain partitioning, (ii) we motivate this new approach with a simple mathematical application model and we show that it always performs at least as good as the standard method of LB, (iii) from this model, we derive an approximation of the optimal LB intervals that can be used for adaptive LB, (iv) we show, using a theoretical and a numerical study, that this new approach can improve the performance of parallel iterative applications.

The paper is organized as follows:
Section~\ref{sec:standard_lb} proposes a formal definition of LB, a review on the previous works, and the limitations of the standard method of LB.
Section~\ref{sec:unloading_lb} introduces our new LB approach that benefits from anticipating the load imbalance growth and the methodology to approximate the optimal LB intervals.
Section~\ref{sec:eval} presents a theoretical analysis using an empirical approach and a numerical study of the proposed method.
Section~\ref{sec:conclusion} concludes the paper and presents our plans for future works.


\section{THE STANDARD LOAD BALANCING METHOD} \label{sec:standard_lb}
Consider an parallel iterative application $A$ consisting of $I$ iterations with an initial workload of $W_{\text{tot}}(0)$ floating point operations.
When dealing with dynamic applications this workload may change from one iteration to the next, hence \hbox{$W_{\text{tot}}(0) \neq W_{\text{tot}}(1) \ ... \neq W_{\text{tot}}(I)$}. 

To compute such an application in parallel on $P$ processing elements (PE), the work units are distributed among the PEs involved in the computation. To obtain optimal performance the workload attributed to each PE must be roughly equal at each iteration.  This is achieved through LB mechanisms. 
In this paper, we are interested in LB mechanisms for parallel iterative applications where workload partitioning is done by domain decomposition. 

\subsection{Formal definition of load balancing}
The LB problem is twofold: (i) find a partitioning of the current workload ($W_{\text{tot}}(i)$) into $P$ partitions, where $P$ is the number of PEs, such that each partition has almost the same workload and (ii) find a mapping of the partitions on the PEs such that the wall clock time to compute those partitions in parallel is minimal~\cite{Pearce2014}. Unfortunately, it has been shown in previous works that the LB problem is fundamentally a balanced graph partitioning problem, which is \hbox{NP-Complete}~\cite{Garey1979}. Hence, finding a good partitioning and mapping is often time consuming. 

Many domain partitioning algorithms have been developed over time achieving good load balance for various applications~\cite{Devine2005, Fleissner2008, Fattebert2012, Begau2015}. However, predicting the time taken by these algorithms to partition and migrate the data (i.e., the cost) is challenging~\cite{Lee2007,Calotoiu2018}.  

A straightforward way to correct load imbalance in an iterative application is to call the load balancer periodically, for instance every $1000$ iterations. However, it is now widely accepted that this method may not adapt to the application requirements~\cite{Pearce2012, H.MenonandN.JainandG.ZhengandL.Kale2012}. So far, the best methodology is to call the load balancer when the LB cost is lower than the cost induced by the load imbalance~\cite{Pearce2012, H.MenonandN.JainandG.ZhengandL.Kale2012}. For that purpose, one must have a model for predicting the cost of a LB call along with a cost model of the application to estimate the gain (resp. the loss) induced by calling (resp. not calling) the load balancer. In this paper, we define the LB process explained above as the \textit{standard load balancing method.} 

\subsection{Related Works} 
\label{RelatedWorks}
Pearce et al.~\cite{Pearce2012} proposed an application load model that is aware of the location of computational elements (e.g., some particles to apply a force on, some mesh sites, etc.) and their dependence. They used their model to estimate the cost of correcting a load imbalance. This model was constructed from multiple simulations of diffusive and global LB techniques performed using: (i) the regression model proposed by Lee et al.~\cite{Lee2007}, (ii) empirically determined parameters (e.g., latency, bandwidth, etc.), and (iii) user/application defined parameters. They used their model to select the most suitable LB technique during application execution and computing the cost and benefit of LB. They employed the simple assumption that a LB step is necessary whenever the LB cost is less than the benefit of LB.

Menon et al.~\cite{H.MenonandN.JainandG.ZhengandL.Kale2012} proposed a simple application model, implemented in \textit{Meta-Balancer}, that uses the same idea as the model of Pearce et al.~\cite{Pearce2012} to call the load balancer automatically. The model supposes that the application is dynamic and thus implies a difference of workload between iterations (i.e., $\Delta_W$). They decomposed $\Delta_W$ in two parts: (i) the workload $m$ that goes to the most loaded PE, and (ii) the workload $a$ that increases the average workload. 
Their approach relies on the \textit{principle of persistence}~\cite{LaxmikantV.Kale2002} to base their decisions on collected data. It states that the changes in the computational loads and the communication schemes tend to persist over time. Menon et al. used collected information about the application performance acquired from the Charm++ runtime system to estimate the average LB cost $C$ during application execution. To approximate the application computing time, they decomposed the iteration time as the average load $l_{\hat{a}}$ plus the extra load $l_{\hat{m}}$ of the most loaded PE. Moreover, the average load increases at rate $\hat{a}$ whereas the extra load (additional to $\hat{a})$ of the most loaded PE increases at rate $\hat{m}$. Also, they assumed that a LB step occurs every $\tau$ iterations. The application computing time is then the sum of all the iteration within all LB intervals. Finally, they used a mathematical formulation to compute analytically the optimal LB interval, which is: \(\tau = \sqrt{2C/\hat{m}}\).


Recently, Zhai et al.~\cite{Zhai2018} enhanced the work of Menon et al.~\cite{H.MenonandN.JainandG.ZhengandL.Kale2012} by improving its adaptive LB capability. They allowed to activate a LB step before the optimal LB interval by incorporating an iterative computation of the performance degradation over time. Besides, they developped three new \textit{ad-hoc} LB techniques for an application of flows with compressible multiphase turbulence called CMT-nek. With these new LB techniques and their adaptive LB approach a performance gain by one order of magnitude is observed compared to the static (i.e., no LB step during application execution) recursive spectral bisection previously used in CMT-nek.

\subsection{Limitations}
The main drawback of the standard LB method is that it aims at balancing the load evenly regardless of how it is growing. In such case, right after a LB step, a new imbalance is immediately recreated. 
As extra work accumulates, this can result in a dramatic waste of computing resources. In other words, to maximize performance we must avoid situations where a minority of PEs is working more than the majority. Therefore, it would be highly beneficial to anticipate such situations.

Let us consider a simple application composed of an initial workload of $W_{\text{tot}}(0)$ FLOP and such that $\Delta_W$ FLOP is added to this workload at each iteration. 
This application is computed on $P$ PEs and each PE has a speed of $\omega$ FLOPS. Besides, the workload is assumed to be balanced at iteration~$0$. The workload at iteration $i$ is:
\begin{equation}\label{eq:app_workload}
    W_{\text{tot}}(i) = W_{\text{tot}}(0) + i\Delta_W.
\end{equation} 

The load imbalance is created by adding $a$~FLOP to every PE and $m$~FLOP to the $N$ most loaded PEs at each iteration; such that \hbox{$\Delta_W = aP + mN$}. This translates to $\hat{a} = a + mN/P$ and $\hat{m} = m(P-N)/P$ in the model of Menon et al.~\cite{H.MenonandN.JainandG.ZhengandL.Kale2012}. The standard LB method consists in distributing the whole workload evenly between the PEs at each LB step. Hence, the computing time of the $t$-th iteration after a LB step $LB_p$ is expressed as:
\begin{equation} \label{eq:StandardModelTpar}
    T_{\text{par}}^{\text{std}}(LB_{p}, t) = \frac{1}{\omega}\big[\frac{W_{\text{tot}}(LB_{p})}{P} + (m+a)t\big],
\end{equation}
where $LB_{p}$ is the previous iteration at which the load balancer has been called. As proposed by Menon et al.~\cite{H.MenonandN.JainandG.ZhengandL.Kale2012}, a call to the load balancer is assumed to lead to a perfect LB. The parallel time of the application between two LB calls $LB_{p}$ and $LB_{n}$ is:
\begin{equation} \label{eq:tpar_within_lb}
     T_ {\text{interval}}(LB_{p}, LB_{n}) = C + \sum_{t=LB_{p}}^{LB_{n}-1} T_{\text{par}}^{\text{std}}(LB_{p}, t-LB_{p}),
\end{equation}
where $C$ is the LB cost in seconds (i.e., the time taken by the LB algorithm to partition and migrate the data). The application's total parallel time is then the sum of the time of all the LB intervals that occur during application execution:
\begin{equation} \label{eq:tpar_total}
     T_{\text{total}} = \sum_{(LB_p, LB_n) \in I}  T_ {\text{interval}}(LB_{p}, LB_{n}),
\end{equation}
where $I$ is the set of the LB intervals. Note that Eq.~(\ref{eq:tpar_total}) is a discrete version of the continuous expression proposed by Menon et al.~\cite{H.MenonandN.JainandG.ZhengandL.Kale2012}.

If we look closely at Eq.~(\ref{eq:StandardModelTpar}), we can see that any extra workload $m$, additional to $a$, collected on any PE (or group of PEs) is creating load imbalance. Moreover, if load imbalance continues to accumulate, this will inevitably produce performance degradation and another LB step will be required. However, if the PEs that receive the extra workload would have been emptied (or underloaded w.r.t. the average workload), then the degradation due to load imbalance would have been mitigated for some iterations. Therefore, in this situation, we need to anticipate how load imbalance is growing to maximize the application performance. 

For that purpose, we propose a novel LB approach that consists in:
\begin{enumerate}
    \item Monitoring continuously the workload increase rate (WIR) of each PE and identifying the $N$ overloading PEs.
    \item Load balancing the workload on all the PEs but giving less workload to the $N$ overloading PEs.
    \item Activating a LB step (i.e., going to step $2$) when the LB cost plus the overhead due to the underloading of the PEs with the highest WIR is lower than the performance degradation due to load imbalance since the last LB step.
\end{enumerate}




\section{THE UNDERLOADING LOAD BALANCING APPROACH - ULBA} \label{sec:unloading_lb}
In this section, we describe the underloading LB approach (ULBA) that adapts the workload distribution such that PEs that are continuously overloading receive a smaller workload than other PEs at the LB steps. This approach relies on the \hbox{\textit{principle of persistence}~\cite{LaxmikantV.Kale2002}} to anticipate the growth of load imbalance instead of just reacting to it.

\begin{figure}[t]
  \centering
  \includegraphics[width=\linewidth]{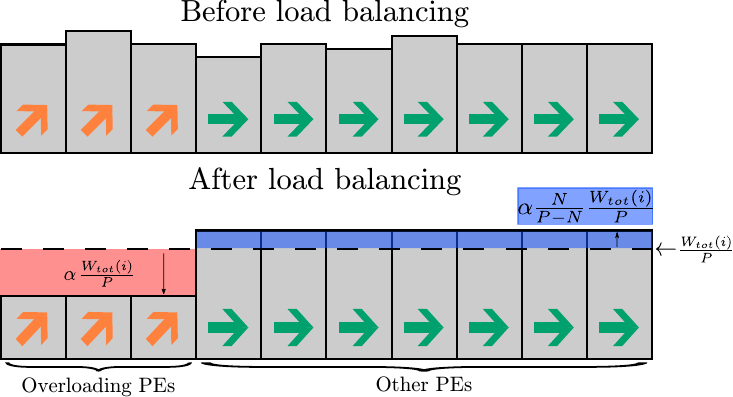}
  \caption{Principle of ULBA. Overloading PEs are marked with an orange arrow while the others are marked with a green arrow. The area colored in red is the total amount of extra workload that overloading PEs will give to the other PEs. The blue area corresponds to the workload gathered by non-overloading PEs from overloading PEs.}
  \label{fig:unloading_model_illustration}
\end{figure}

Figure~\ref{fig:unloading_model_illustration} illustrates how ULBA works. At a LB step, the load balancer unloads each overloading PE by a fraction $\alpha$ of the perfectly balanced workload (red area). The other PEs then receive the sum of this extra workload fairly divided among them (blue area). The formula inside the red area describes amount of workload transferred by a single overloading PE, whereas the formula inside the blue area is the amount of workload gathered by a single non-overloading PE. 

\subsection{Formal definition of ULBA}
In this section, we formally describe the behavior of ULBA as well as its parameters.

ULBA employs monitoring and anticipation mechanisms to adapt the workload distribution to the application dynamics. 
To model the theoretical parallel computing time, we use application parameters that are similar to those in the work of Menon et al.~\cite{H.MenonandN.JainandG.ZhengandL.Kale2012} (Table~\ref{table:MyModelParameters}). 
The workload at a particular iteration is defined in Eq.~(\ref{eq:app_workload}). 
At each LB call, the PEs that have the highest WIR transfer a fraction $\alpha$ of the perfectly balanced workload to the PEs that do not overload. The WIR of the $N$ overloading PEs is $m+a$, whereas the WIR of the other PEs is $a$. If there is no overloading PE (i.e., $m = 0$) then there is no reason to use ULBA as there is no load imbalance creation over time.

With ULBA, the time to compute an iteration right after a LB call is dominated by the $P-N$ non-overloading PEs. Once the overloading PEs have reached the same amount of work as the other ones, the following iterations costs are dominated by the overloading PEs that continue to accumulate extra workloads. When a call to the load balancer occurs (e.g., at iteration $i$), \(\alpha W_{\text{tot}}(i)/P\) work units are transferred from each of the $N$ overloading PEs to the $P-N$ non-overloading PEs. Hence, the parallel time to compute the $t$-th iteration after the LB step $LB_p$ is now expressed as:
\begin{equation} \label{eq:NewModelTpar} 
	\resizebox{0.5\textwidth}{!}{$T_{\text{par}}^{\text{ULBA}}(LB_{p}, t) = \frac{1}{\omega} 
    \begin{cases}
       (1+\frac{\alpha N}{P-N})\frac{W_{\text{tot}}(LB_{p})}{P} + at, & \hspace{-0.5pt}\text{if} \ t \leq \sigma^{\tiny-}(LB_p) \\
       (1-\alpha) \frac{W_{\text{tot}}(LB_{p})}{P} + (m+a)t, & \text{otherwise,} \\
    \end{cases}$}
\end{equation}
where $\sigma^{\tiny-}(LB_{p})$ is the number of iterations needed for the overloading PEs to accumulate the same load as the others (Eq.~(\ref{eq:lowerbound2})). The application parallel time is computed as in the standard method of LB by using Eq.~(\ref{eq:tpar_total}) except that we substitute Eq.~(\ref{eq:StandardModelTpar}) by Eq.~(\ref{eq:NewModelTpar}) in Eq.~(\ref{eq:tpar_within_lb}). In the present paper, we consider that $\alpha$ is constant and user defined for all overloading PEs. However, to take the most out of ULBA, $\alpha$ should depend on the WIR of each overloading PE. The dynamic adaptation of $\alpha$ is left for future works. 
 
\begin{table}[]
\caption{Parameters involved in ULBA}
\makebox[\linewidth] {
\begin{tabular}{ll}
\hline\hline
Name         & Description\\
\hline
$P$          & Number of PEs.\\ 
$N$          & Number of overloading PEs.\\
$\gamma$     & Number of iterations during which the application runs.\\
$W_{\text{tot}}(i)$ & Workload at iteration $i$; $W_{\text{tot}}(0) = $ initial workload.\\ 

$\hat{a}$    & Average workload increase rate~\cite{H.MenonandN.JainandG.ZhengandL.Kale2012}.\\
$\hat{m}$    & Workload increase rate (additional to $\hat{a}$) of the most loaded PEs~\cite{H.MenonandN.JainandG.ZhengandL.Kale2012}.\\ 
$a$          & Amount of workload that goes to every PE at each iteration.\\
$m$          & Workload additional to $a$ that goes to the overloading PEs.\\ 

$\Delta_W$   & Workload difference between two iterations; $\Delta_W = aP + mN$. \\ 
$\alpha$     & Fraction of workload to remove from overloading PEs. \\
$\omega$     & Speed of every PE.\\
$C$          & Cost of performing a LB step. \\  
$LB_{p}$     & Iteration of the previous LB call.\\  
$LB_{n}$     & Iteration of the next LB call.\\  
$I$          & The set of all the LB intervals.
\end{tabular}
}
\label{table:MyModelParameters}
\end{table}

\subsection{Finding the optimal load balancing intervals}
In ULBA, there exists, after each LB step, an optimal number of iterations to wait before calling the load balancer again. However, as early decisions influence future LB calls, finding the optimal LB intervals is challenging using an analytical method.
Instead, we propose a range $[\sigma^{\tiny-}$, $\sigma^{\tiny+}]$ of iterations within which the next LB call should occur to obtain a performance close to the optimal LB interval. 

As stated above, at a LB step a fraction $\alpha$ of the perfectly balanced workload of each overloading PE is transferred to the $P-N$ non-overloading PEs. Let $W^*$ be the workload of any overloading PE and $W$ the workload of any non-overloading PE right after a LB step. According to Figure~\ref{fig:unloading_model_illustration}, they are expressed as:
\begin{equation}\label{eq:workload_after_lb}
\small \begin{split}
        W^* &= (1-\alpha) \frac{W_{\text{tot}}(LB_{p})}{P},\\ 
        W &= (1+\frac{\alpha N}{P-N})\frac{W_{\text{tot}}(LB_{p})}{P}.
\end{split}
\end{equation} 
Based on their WIR, overloading PEs will take some iterations to reach $W$. Let call this number of iteration $\sigma^{\tiny-}$, which we refer to as the LB lower bound. Mathematically, this is given by the height of the red area plus the height of the blue area (see Figure~\ref{fig:unloading_model_illustration}) divided by the difference of WIR between overloading PEs and non-overloading PEs. Hence, after $\sigma^{\tiny-}$ iterations since the last LB step at iteration $i$ the workload of overloading and non-overloading PEs must be the same, leading to the following equations:
\begin{equation} \label{eq:lowerbound1}
    \sigma^{\tiny-}(i) (m+a) + W^* = \sigma^{\tiny-}(i)a + W.
\end{equation}
By substituting $W^*$ and $W$ by their respective values (see Eq.~(\ref{eq:workload_after_lb})) in Eq.~(\ref{eq:lowerbound1}), we can derive the LB lower bound:
\begin{equation} \label{eq:lowerbound2}
\resizebox{0.5\textwidth}{!}{$\begin{split}
    \sigma^{\tiny-}(i) (m+a) + (1-\alpha) \frac{W_{\text{tot}}(i)}{P} &= \sigma^{\tiny-}(i)a + (1+\frac{\alpha N}{P-N})\frac{W_{\text{tot}}(i)}{P},\\
    \sigma^{\tiny-}(i)m &= \alpha \frac{W_{\text{tot}}(i)}{P} + \frac{\alpha N}{P-N}\frac{W_{\text{tot}}(i)}{P},\\
    \sigma^{\tiny-}(i) &= \Bigl\lfloor\big(1+\frac{N}{P-N}\big)\frac{\alpha W_{\text{tot}}(i)}{m P}\Bigr\rfloor.
\end{split}$}
\end{equation}

After a call to the load balancer at iteration $i$ there is no gain to recall the load balancer before $\sigma^{\tiny-}(i)$ because there is no degradation over time due to load imbalance until the overloading PEs reach the critical load $W^* = W$. 

The upper bound of the number of iteration to wait before calling the load balancer can be understood as the largest $i$ such that a later call leads to worse performance. 
To find the upper bound we decided to use the same procedure as proposed by Menon et al.~\cite{H.MenonandN.JainandG.ZhengandL.Kale2012}. A LB step is activated when the load imbalance cost is equal to the average LB cost $C$ (in seconds). In their approach there is no overhead due to the workload distribution because it leads to a perfect load balancing. Hence, their LB interval $\tau$ is such that~\cite{H.MenonandN.JainandG.ZhengandL.Kale2012} \[ \text{Cost}_{\text{imbalance}}(\tau) = C.\] Consequently, we decided to use the same equation to compute the upper bound except that we also take into account the overhead of ULBA:
\begin{equation} \label{eq:upper_bound_equation}
    \text{Cost}_{\text{imbalance}}(\tau) = \text{Cost}_{\text{overhead}}(LB_p, \tau) + C.
\end{equation}
The load imbalance cost over $\tau$ iterations since $\sigma^{\tiny-}(LB_p)$ is defined as the sum of the workload increase rate, additional to the average workload increase rate, of the most loaded PEs (i.e., $\hat{m}$)~\cite{H.MenonandN.JainandG.ZhengandL.Kale2012}:
\begin{equation} \label{eq:cost_imbalance}
    \text{Cost}_{\text{imbalance}}(\tau) =  \frac{1}{\omega} \int_{0}^{\tau} \hat{m}t \diff t.
\end{equation}
Changing the integral into a discrete sum only leads to a non-significant change in the value of the upper bound.

The overhead created by ULBA is proportional to the amount of workload transferred to non-overloading PEs. It is the amount of workload a single non-overloading PE (the height of the blue area in Figure~\ref{fig:unloading_model_illustration}) will get from the overloading PEs at the next LB step (i.e., at step $LB_p+\sigma^{\tiny-}(LB_p)+ \tau$): 
\begin{equation} \label{eq:cost_overhead}
    \text{Cost}_{\text{overhead}}(LB_p, \tau) = \frac{\alpha N}{P-N} \frac{W_{\text{tot}}(LB_p+\sigma^{\tiny-}(LB_p)+ \tau)}{\omega P}.
\end{equation}

Substituting, solving, and rearranging the terms in Eq.~(\ref{eq:upper_bound_equation}) leads to the following quadratic equation:
\begin{equation} \label{eq:quadratic_equals_lb2} 
\resizebox{0.5\textwidth}{!}{$\frac{\hat{m}}{2\omega} \tau^2 - \frac{\alpha N \Delta_W}{(P-N)\omega P}\tau - \big[\frac{\alpha N}{P-N} \frac{W_{\text{tot}}(LB_p) + \sigma^{\tiny-}(LB_p)\Delta_W }{\omega P} + C\big]= 0.$}
\end{equation}
Solving this equation leads to two points: $(\tau_1, \tau_2)$ from which we define $\sigma^{\tiny+}(i) = \sigma^{\tiny-}(i) + \max (\tau_1, \tau_2)$. Note that, the proposed approach behaves like the standard LB method when $\alpha$ is set to zero. In this case, $\sigma^{\tiny-}(i)~=~0$ and \(\sigma^{\tiny+}(i)~=~\sqrt{2C/\hat{m}}\). 

\begin{table}[]
\caption{Random application parameters distribution.}
\makebox[\linewidth] {
\begin{tabular}{ll}
\hline\hline
Name         & Distribution\\
\hline
$P$          & Uniformly sampled on [$256$, $512$, $1024$, $2048$]\\ 
$N$          & $P\cdot v, \ v \sim $ Uniform($0.01$, $0.2$)      \\
$\gamma$     & $100$\\
$W_{\text{tot}}(0)$ & Uniform($52 \cdot 10^7 \cdot P$, $1165 \cdot 10^7 \cdot P$) \\ 
$\Delta_W$   & $\frac{W_{\text{tot}}(0)}{P} \cdot x, \ x \sim $ Uniform($0.01$, $0.3$) \\
$a$          & $\frac{\Delta_W}{P} \cdot (1-y), \ y \sim $ Uniform($0.8$, $1.0$) \\
$m$          & $\frac{\Delta_W}{N} \cdot y$\\ 
$\alpha$     & $\alpha \sim $ Uniform($0.0$, $1.0$) \\
$C$          & $\frac{W_{\text{tot}}(0)}{P} \cdot z, \ z \sim $Uniform($0.1$, $3.0$)
\end{tabular}


}
\label{table:HeuristicRandomParameters}

\end{table}

To validate the lower and upper bounds, we used an heuristic search approach (simulated annealing) to find an approximation of the optimal LB intervals for a given application instance. We verified that a solution obtained using simulated annealing is close (in term of theoretical performance) to load balancing the application at each $\sigma^{\tiny+}$. A state is a vector of booleans of size $\gamma$ that contains the LB state of each iteration. A value ``true'' (resp. ``false'') means that we call (resp. not call) the load balancer. The heuristic search algorithm can move inside the state space by activating or deactivating the load balancer at a particular iteration. The cost function to minimize is Eq.~(\ref{eq:tpar_total}) using Eq.~(\ref{eq:NewModelTpar}) in Eq.~(\ref{eq:tpar_within_lb}) (i.e., summing all the step time for all the LB intervals). 

We compared the performance of the sequence of LB intervals found by the simulated annealing with a LB step every $\sigma^{\tiny+}$ iterations. We performed this comparison on a set of $1000$~different application instances. For each simulation, we draw a random value for each parameter in Table~\ref{table:HeuristicRandomParameters}. We fixed the application duration (i.e., $\gamma$) to $100$ iterations and, for the sake of simplicity, the speed of PE (i.e., $\omega$) to one GFLOPS. We tried to have a large enough input space to show that the upper bound gives a good approximation of the optimal LB intervals.  Moreover, the workload bounds are set to be in the magnitude of a 2D/3D computational fluid dynamic application involving $10^7$ cells per process. The lower bound is $52$ FLOP per cells, whereas the upper bound is $1165$ FLOP per cells~\cite{tomczak2018sparse}. The number of overloading PEs is at most $20\%$ of the total number of PEs. The application WIR is between $1\%$ and $10\%$ of the workload per PE. At each iteration, $20\%$ percent of the application WIR is distributed to every PE and $80\%$ percent is distributed, in addition, to the overloading PEs in order to consider only imbalanced applications.
The average LB cost ranges between $10\%$ to $100\%$ of the time to compute one iteration. Each result obtained with the heuristic search took, in average, $120$~seconds to converge to the optimal LB intervals on Intel i7-8700 CPU ($3.20$~GHz) with the python ``simanneal'' module~\cite{PySimanneal}.

\begin{figure}[ht]
  \centering
  \includegraphics[width=\linewidth]{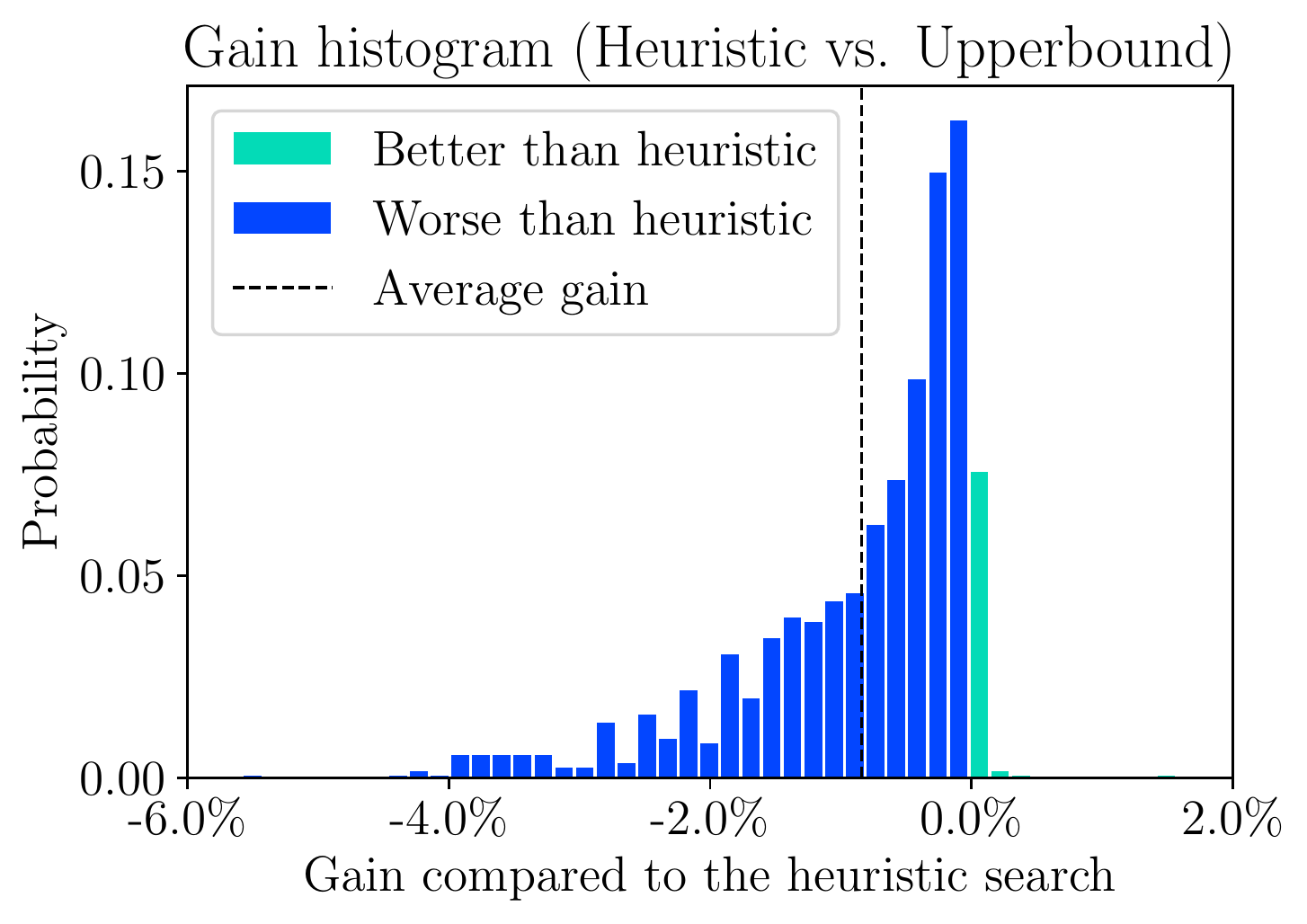}
  \caption{Probability distribution of gains on $1000$~different application instances between the LB intervals found via the heuristic search (simulated annealing) and the LB intervals found using the upper bound.}
  \label{fig:heuristic_vs_upperbound}
\end{figure} 

Figure~\ref{fig:heuristic_vs_upperbound} presents the histogram of the relative difference of wall clock time between the simulated annealing optimum and $\sigma^{\tiny{+}}$. It shows that the proposed approach is always very close to the heuristic solution sampled over $1000$~instances. The better gain is $+1.57\%$ of the total solution cost, while the worse gain is $-5.58\%$ and the average gain is $-0.83\%$. This demonstrates that, while not being optimal, the analytical approach is a good approximation of a solution found via a numerical optimization. Therefore, as the range $[\sigma^{\tiny-}, \sigma^{\tiny+}]$ can be wide, we propose to use $\sigma^{\tiny+}$ as the LB steps.

\subsection{Implementation of ULBA} \label{sec:impl_ulba}
In this section, we deal with the implementation of ULBA with a centralized LB technique. To recall, at a LB step, ULBA gives less work to the PEs that are predicted to be overloaded in the near future. To know whether a PE is overloading, each PE keeps a database that stores the WIR of every PE. Each PE evaluates its WIR and propagates it (as well as the most recent WIRs in its database) to the other PEs using a dissemination algorithm (or gossip algorithm~\cite{Demers1987,Amar2009}). Herein, one dissemination step is done at each iteration to mitigate the overhead due to the WIR communication. Indeed, the WIRs received from other PEs are still ``up to date'' few steps after receiving them when assuming a \textit{principle of persistence}~\cite{LaxmikantV.Kale2002}. 
 
\alglanguage{pseudocode}
\begin{algorithm}
\caption{ULBA: Application Skeleton}\label{prepare_lb_algo}
\begin{algorithmic}[1]
\State MAX\_STEP $ \gets $ Number of steps to do
\State data $\gets$ Work units to compute \Comment{The workload is initially balanced}
\State $\alpha \gets$ Extra amount to transfer if I am overloading
\State threshold $\gets$ Outlier threshold
\State rank $\gets$ My MPI rank
\State irdb $ \gets $IncreaseRateDatabase
\State lb\_step $\gets 0$
\State degradation $\gets 0$
\For {$i \rightarrow 0..$MAX\_STEP}
    \State time$_i$ $=$ \Call{ComputeStep}{$i$, data}\Comment{Data movements and computation}
    \If{i == lb\_step}
        \State ref\_time = time$_i$
    \EndIf{}
    \State t = median of time per step in [time$_i$, time$_{i-2}$]
    \State degradation $ = $ degradation $+$ $($t $-$  ref\_time$)$
    \If{degradation $ \geq $ avg\_lb\_cost}
        \State $s = $ GetIncreaseRateOf(irdb, rank)
        \State $S = $ GetAllIncreaseRates(irdb)
        \If{z-score($s, S$) $< $ threshold}\Comment{I am not overloading}
            \State \Call{CallLoadBalancer}{i, rank, data, $0.0$}
        \Else
            \State \Call{CallLoadBalancer}{i, rank, data, $\alpha$}
        \EndIf{}
        \State lb\_step $ = i+1$
        \State degradation $=$ 0
    \EndIf{}
\EndFor 

\end{algorithmic}
\end{algorithm}

\alglanguage{pseudocode}
\begin{algorithm}
\caption{ULBA: Example with a centralized load balancer}\label{call_lb_algo_centralized}
\begin{algorithmic}[1]
\Function{CallLoadBalancer}{$i$, rank, data, $\alpha$}
\State $W_{\text{tot}}\text{(i)} \gets $ Total Workload at iteration~$i$
\State $P \gets $ Number of PEs
\State \Call{SendAlphaToMainPE}{$\alpha$}
\If{MainPE == rank}
    \State $A = $ \Call{RecvAlphas}{}
    \State $N = $ \Call{GetNumberOfOverloadingPEs}{A}
    \For {$p \rightarrow 1..P$}
        \If{$A_p > 0.0$}\Comment{$p$ is overloading}
            \State w$_p = (1-A_p)W_{\text{tot}}\text{(i)}\frac{1}{P}$
        \Else
            \State w$_p = \big(1+\frac{A_p\cdot N}{P-N}\big)W_{\text{tot}}\text{(i)}\frac{1}{P}$
        \EndIf{}
    \EndFor{}
    \State \Call{PartitionAccordingToWeights}{w}
    \State \Call{BroadcastPartition}{}
\Else
    \State \Call{ReceivePartition}{}
\EndIf{}
    \State \Call{MigrateDataAccordingToPartition}{data}
\EndFunction{}
\end{algorithmic}
\end{algorithm}

As proposed in the previous section, the load balancer is called every time the degradation due to load imbalance overcomes the average LB cost plus the overhead of ULBA. We implemented this behavior using the approach proposed by Zhai et al.~\cite{Zhai2018} that computes the exact degradation of each iteration w.r.t. a reference iteration (in our case, the one just after the last LB call). At a LB step, each PE detects whether it is an overloading PE or not. A PE is considered overloading if the \textit{z-score} of its WIR in the distribution of the WIR created from the database exceeds $3.0$. As soon as a PE detects that it overloads, it sets $\alpha$ (see Table~\ref{table:MyModelParameters}) to a user defined value in the range $[0, 1]$, otherwise it sets it to $0$. Afterwards, all the PEs call the load balancer with their respective $\alpha$, meaning that they will keep a fraction ($1-\alpha$) of $W_{\text{tot}}(i)/P$ (see Eq.~(\ref{eq:workload_after_lb})).
If at least $50\%$ of the PEs call the load balancer with $\alpha > 0$, then, the load balancer works as the standard LB method (i.e., $\alpha = 0$ for all the PEs) because it is counter-productive to unload a majority of PEs. An application skeleton for ULBA is proposed in Algorithm~\ref{prepare_lb_algo}. An implementation of the function ``CallLoadBalancer'' is given as an example for a centralized LB technique in Algorithm~\ref{call_lb_algo_centralized} and it is used in the centralized LB technique presented in the following section. This function partitions the work according to the value $\alpha$ of each PE and migrates the workload accordingly. We plan to implement ULBA for other types of LB technique in a widely used LB framework (e.g., the Zoltan framework~\cite{Devine2002}) in future works.

\section{EXPERIMENTAL EVALUATION} \label{sec:eval}
In this section, we present two evaluations of ULBA:
\begin{enumerate}
    \item An empirical analysis of the performance gain obtained from using ULBA compared to the standard LB method. For ULBA, we used Eq.~(\ref{eq:tpar_total}), with Eq.~(\ref{eq:NewModelTpar}) in Eq.~(\ref{eq:tpar_within_lb}), to get the theoretical computing time. For the standard LB method, we employed Eq.~(\ref{eq:tpar_total}), with Eq.~(\ref{eq:StandardModelTpar}) in Eq.~(\ref{eq:tpar_within_lb}), to compute the theoretical computing time.
    \item A numerical study using an application that reproduces the computation of a fluid model with non-uniform erosion. In this application, we implemented a centralized LB technique on which we applied the adaptive LB approach of Zhai et al.~\cite{Zhai2018} and ULBA. We compared the computation time and the average PE usage of these two approaches.
\end{enumerate}
\subsection{Simulations}
As a first evaluation, we compared the performance of the standard LB method and ULBA w.r.t. the percentage of overloading PEs ($N/P$). We varied the percentage of overloading PEs from $1.0\%$ to $20\%$. For each percentage of overloading PEs, we compared the performance of the two approaches on $1000$~different application instances. For ULBA and for each application instance, we tested $100$ values of $\alpha$ uniformly distributed in the range [$0$, $1$] and we kept the value that maximizes the performance. We randomly sampled the application parameters from Table~\ref{table:HeuristicRandomParameters} (except $P$, $N$, and $\alpha$). We used these parameters in Eq.~(\ref{eq:tpar_total}), with Eq.~(\ref{eq:StandardModelTpar}) in Eq.~(\ref{eq:tpar_within_lb}), to get the parallel computation time of the standard LB method. For ULBA, we used the same application instances and we utilized Eq.~(\ref{eq:tpar_total}), with Eq.~(\ref{eq:NewModelTpar}) in Eq.~(\ref{eq:tpar_within_lb}), to compute the parallel computation time. Then, we analyzed the percentage of gain we obtain by using ULBA in comparison with the standard LB method. 

\begin{figure}[t]
  \centering
  \includegraphics[width=\linewidth]{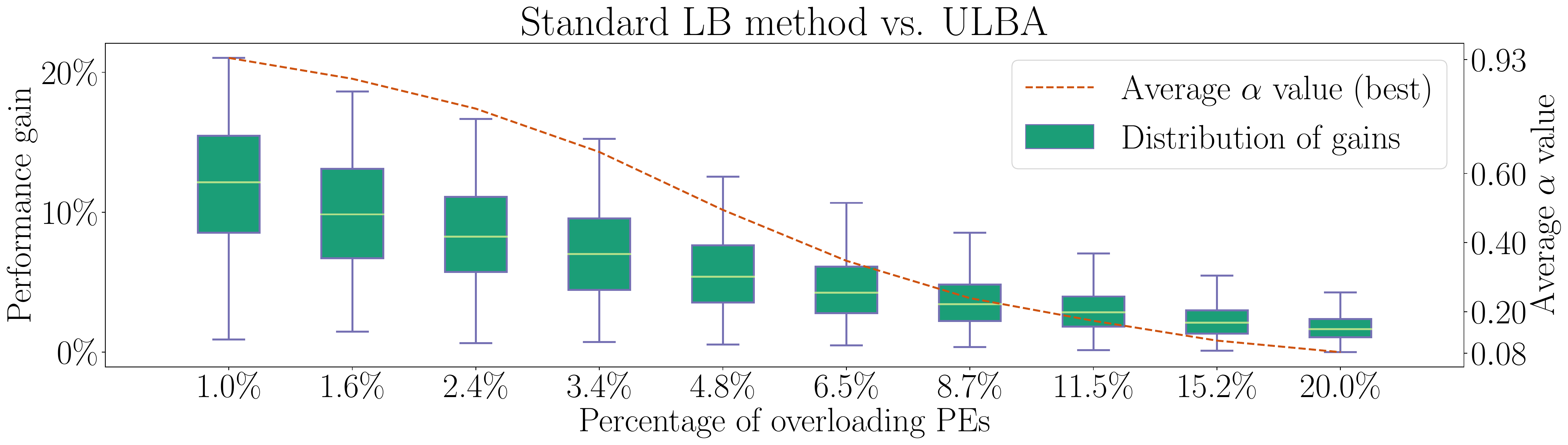}
  \caption{Box plots of empirical analysis of the theoretical performance gain between the standard LB method and ULBA w.r.t. the percentage of overloading PEs. ULBA can increase the performance by up to $21\%$ compared to the standard LB method.}
  \label{fig:unloading_vs_menon_simulations}
\end{figure}

Figure~\ref{fig:unloading_vs_menon_simulations} shows the performance comparison of the standard method of LB with a discrete version of the model of Menon et al.~\cite{H.MenonandN.JainandG.ZhengandL.Kale2012} and ULBA w.r.t. the percentage of overloading PEs. Besides, this figure shows the average value of $\alpha$ used in the application instances w.r.t. the percentage of overloading PEs. First, it illustrates that ULBA is always better than the standard LB method. As shown in Section~\ref{sec:unloading_lb}, ULBA works as the standard LB method when $\alpha = 0$, hence, there is always a $\alpha$ that performs at least as good as the standard method of LB. On the other hand, Figure~\ref{fig:unloading_vs_menon_simulations} shows that ULBA can increases the performance of an application by up to $21\%$ compared to the standard method of LB. However, it performs best when the percentage of overloading PEs is low. Indeed, the overhead due to ULBA grows linearly as function of the ratio of overloading PEs and $\alpha$. Therefore, to take the most out of ULBA, $\alpha$ should be as high as possible, however, it must also decrease as the percentage of overloading PEs increases. This is a consequence of the overhead due to the total amount of workload transferred from overloading PEs to non-overloading PEs. As the percentage of PEs that continuously overload can change during application execution, $\alpha$ must be adapted at runtime, nevertheless, we let this task for future works. 

\subsection{Using a numerical study}
To evaluate ULBA, we decided to compare the performance of an imbalanced dynamic parallel application with the standard method of LB that uses the adaptive LB approach of~\cite{Zhai2018} and ULBA. 

This parallel application reproduces the computation of a fluid and the erosion of immersed rocks. The size of the computational domain, the number of rocks and their size are input parameters of the application. The computational domain is organized as a 2D mesh with two cell types: fluid and rock. Each fluid cell computes a probabilistic erosion of neighboring rock cells and reproduces the computation of a fluid, whereas rock cells involve no computation. A rock is an aggregate of rock cells organized in a disc, the rock cells within a given disc share the same erosion probability. To create imbalance in the computational effort needed in different parts of the domain, the erosion probability attached to a rock is either $0.02$ or $0.4$. It is not known in advance where the rocks with a high eroding probability are located. Moreover, when a fluid cell erodes a rock cell, it converts the rock cell into four fluid cells of smaller size reproducing a mesh-refinement mechanism, creating even more imbalance.  
To distribute the application workload, we implemented a partitioning technique that divide the computational domain in stripes along the x-axis. Consequently, a stripe is composed of several consecutive columns of cells. The goal of this technique is to create $P$ stripes that roughly contains the same number of fluid cells. This technique is implemented as a centralized LB technique where the stripe associated to each PE is computed on a single PE and then broadcasted to the others. This LB technique implements ULBA and enables adaptive LB using the methodology presented in Section~\ref{sec:impl_ulba}. Each processing element deals with the cells within its stripe. Hence, the stripes that contain rock cells that have a higher probability of being eroded will gather more workload than other stripes as the simulation runs. 

We performed two experimental analyses on Baobab, the cluster of the University of Geneva on \hbox{\textit{Intel Sandy Bridge-EP E5-2630V4}} ($2.2$ Ghz):

\paragraph{Performance comparison} We compared the performance of the standard method of LB with the adaptive LB approach proposed by Zhai et al.~\cite{Zhai2018} against ULBA. We fixed the size of the domain to \hbox{$(P\cdot1000)$x$1000$}~cells (i.e., $1$~million cells per PE). $P$ rock disks with a radius of $250$~cells are uniformly distributed along the x-axis. At the beginning of the application, the partitioning technique attributes one rock per PE. No PE knows whether the rock disk located in its stripe is strongly or weakly erodible. We scaled the number of PEs from $32$ to $256$. For each method, we compared the median running time among five runs. Note that, we executed ULBA with $\alpha = 0.4$.

\begin{figure}[t]
  \centering
  \includegraphics[width=\linewidth]{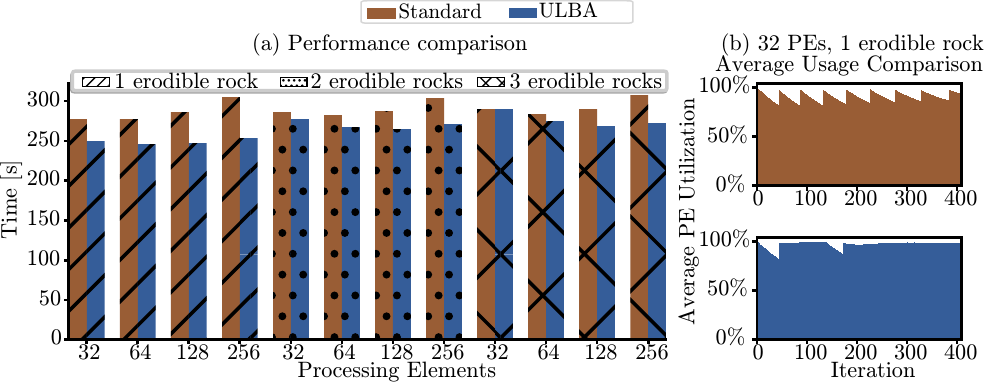}
  \caption{Performance comparison between the standard LB method with the adaptive LB approach of Zhai et al.~\cite{Zhai2018} and ULBA on an application that reproduces a fluid model with the erosion of $1$ to $3$ strongly erodible rocks among a total of $P$ rocks. Each rock contains thousands of rock cells. The implementation of the standard LB method and ULBA use the same centralized LB technique. The performance of ULBA is colored in blue, while the performance of the standard LB method is colored in brown. The results show an higher average PE usage for ULBA as well as a performance improvement up to $16\%$.}
  \label{fig:perf1}
\end{figure}

Figure~\ref{fig:perf1}a shows the performance comparison between the standard LB method with the adaptive LB approach of Zhai et al.~\cite{Zhai2018} and ULBA. Figure~\ref{fig:perf1}b shows the average PE usage (in percent) on one test case involving $32$~PEs and one strongly erodible rock.

From Figure~\ref{fig:perf1}a, we observe that ULBA performs better than the standard LB method on all the test cases except for one where the performance is equal. As pointed out earlier, the performance of ULBA decreases as the percentage of overloading PEs increases. For example, in Figure~\ref{fig:perf1}a, the performance of the cases with $32$~PEs decreases as the number of strongly erodible rocks increases. Hence, for the case with the greater percentage of overloading PEs (i.e., $32$~PEs with $3$~strongly erodible rocks), the performance is equal to the standard method of LB. Moreover, Figure~\ref{fig:perf1}a illustrates that, for a given number of erodible rocks, ULBA scales well with the number of PEs. 
Furthermore, we see in Figure~\ref{fig:perf1}b that ULBA is more capable of withstanding to the load imbalance growth as shown by less drops in the CPU usage. We also observe that ULBA increases the average PE usage. Consequently, Figure~\ref{fig:perf1}b shows that ULBA involves less call to the LB technique ($62.5\%$ fewer LB calls). Note that ULBA wasted a LB call at iteration~$315$ (Figure~\ref{fig:perf1}b). 

\paragraph{Tuning of the $\alpha$ parameter} During the previous experiments we identified that changing $\alpha$ leads to significant performance variations. Hence, we decided to study how this parameter affects the performance of ULBA. We compared the performance of ULBA with various $\alpha$ for the application presented above. We kept the same domain size and rock disks size, however, we studied only the case with one strongly erodible rock among $P$ rocks. We executed the application on $32$, $64$, $128$, and $256$ PEs. 
\begin{figure}[t]
  \centering
  \includegraphics[width=0.78\linewidth]{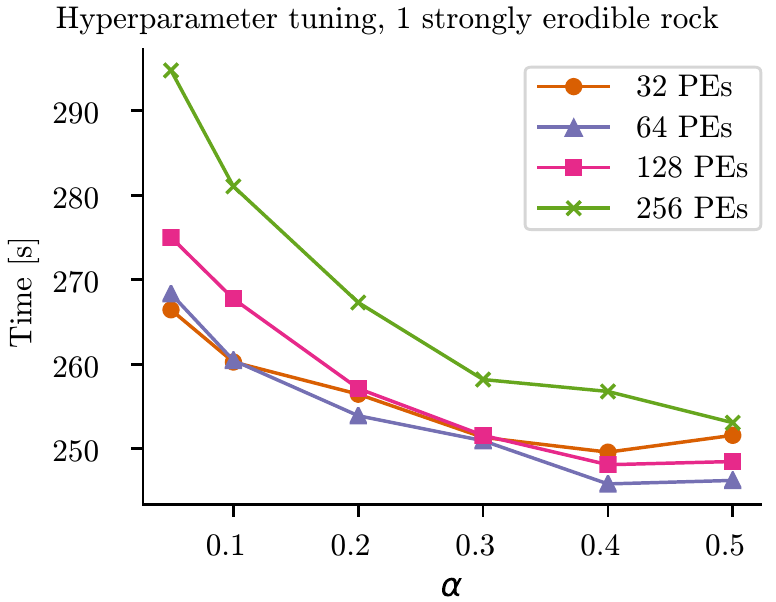}
  \caption{Performance of ULBA w.r.t. $\alpha$ on an application that reproduces a fluid model with erosion. The value of $\alpha$ strongly impacts the performance of ULBA. There is no significant gain to increase the value of $\alpha$ above $0.4$ ($32$, $64$, $128$ PEs) except for $256$ PEs that observes a performance improvement of $1.4\%$ by increasing $\alpha$ from $0.4$ to $0.5$.}
  \label{fig:alpha1}
\end{figure}
Figure~\ref{fig:alpha1} shows the performance of the application w.r.t. the value of $\alpha$. First, it shows that $\alpha$ has a strong impact on the application performance (up to $14\%$ performance gain). On the other hand, we observe that there is no significant gain to increase $\alpha$ above $0.4$ for $32$, $64$, and $128$ PEs. However, it still increases the performance for $256$ PEs. As shown in Eq.~(\ref{eq:cost_overhead}), the overhead of ULBA depends linearly on $(\alpha N)/(P-N)$. This suggests that, for a given overhead, $\alpha$ can be set higher whether $N/(P-N)$ is small, resulting in better overall performance. We also observed in Figure~\ref{fig:unloading_vs_menon_simulations} that greater $\alpha$ values were associated with better performance gains. 

These results validates the benefits of anticipating the load imbalance to improve the performance of parallel applications. It also indicates that the additional parameter $\alpha$ should be determined during application execution because it depends on variable parameters such as the number of overloading PEs. Moreover, defining the value that $\alpha$ should take to maximize application performance is still an open question. Hence, we plan to identify in a further work the exact value that $\alpha$ must take to maximize the application performance and the adjustment of its value at runtime.



\section{CONCLUSION} \label{sec:conclusion}
In the present paper, we have presented a first step toward an anticipation-based LB approach that, unlike the standard LB method, gives less workload to processing elements that are continuously overloading in order to optimize application performance. 
We presented a simple mathematical application model and we used it to find an approximation of the optimal LB intervals, enabling adaptive LB. We showed that this method was not far from a solution found via a numerical optimization. This analytical solution balances the application load automatically requiring only a measure of the performance degradation since the last LB step, an estimation of the ULBA overhead, and the average LB cost. 

In addition, we implemented the proposed approach with a centralized LB technique and we applied it on an application that reproduces a fluid model with non-uniform erosion. We compared the performance obtained with our approach to the performance of standard method of LB with the adaptive LB approach proposed by Zhai et al.~\cite{Zhai2018}. We observed that the proposed approach improved the performance of our application by up to $16\%$ validating the benefits of anticipating the load imbalance growth.

As the proposed approach involves a new parameter (i.e., $\alpha$), we performed a study of the impact of this parameter on the application performance. We showed that the value of $\alpha$ impacts strongly the performance of the proposed approach. Besides, we identified that $\alpha$ depends on the ratio of overloading PEs and thus should be adapted during application execution. 

We plan to integrate our approach in a widely used load balancing suite such as Zoltan~\cite{Devine2002}, to define the value that $\alpha$ should take to optimize the application performance, and to dynamically adjust $\alpha$ during application execution in future works.  

\section{ACKNOWLEDGEMENTS}
This work has benefited from the support of the CADMOS project.

\bibliographystyle{./bibliography/IEEEtran}
\bibliography{ref}

\end{document}